\documentclass[a4paper,12pt]{article}

\usepackage{ifpdf}

\newif\ifpdf
\ifx\pdfoutput\undefined
  \pdffalse
\else
  \pdfoutput=1
  \pdftrue
\fi

\RequirePackage{xspace} %
\RequirePackage{subfigure} %
\RequirePackage[centertags]{amsmath} %
\RequirePackage{amssymb}
\RequirePackage{wrapfig} %
\RequirePackage{calc} %
\RequirePackage{ifthen}
\RequirePackage{tabularx} %
\RequirePackage{flafter} %
\RequirePackage{fancyhdr} %

\ifpdf
  \RequirePackage[pdftex]{color}%
  \RequirePackage{colortbl}%
  \RequirePackage{array}%
  \RequirePackage[pdftex]{graphicx}

  \RequirePackage[ pdftex, plainpages = false, pdfpagelabels,
                 pdfpagelayout = useoutlines,
                 bookmarks,
                 breaklinks = true,
                 linktocpage,
                 pagebackref,                      
                 colorlinks = true,
                 linkcolor = blue,
                 urlcolor  = blue,
                 citecolor = blue,
                 anchorcolor = blue,
                 hyperindex = true,
                 hyperfigures
                 ]{hyperref}

\else
  \RequirePackage{color}
  \RequirePackage{colortbl}
   \RequirePackage{array}
  \RequirePackage[dvips]{graphicx}
  \RequirePackage{hyperref}
  \usepackage{rotating}
\fi

\usepackage{makeidx} 
\usepackage{setspace} 
\usepackage{rotating} 
\usepackage{ecltree}
\usepackage{epic}
\usepackage{supertabular}  
\usepackage{color}
\usepackage{exscale}
\usepackage{fontenc}
\usepackage{ifthen}
\usepackage{latexsym}
\usepackage{makeidx}
\usepackage{syntonly}
\usepackage{inputenc}
\usepackage{graphicx}
\usepackage{setspace}
\usepackage{caption2}
\usepackage[english]{babel}
\usepackage[square, comma,numbers,sort&compress]{natbib}
\usepackage{hypernat}
\usepackage{boxedminipage}
\usepackage{framed}
\usepackage{longtable}
\usepackage[all]{hypcap}    
\usepackage{algorithm2e}
\usepackage{algorithmic}
\usepackage{lscape}
\usepackage{pdflscape}

\newcommand{\note}[1]{\marginpar[left]{\singlespace \tiny #1}}

\renewcommand{\sectionmark}[1]%
      {\markright{\thesection\ #1}} 

\renewcommand{\note}[1]{}

\newcommand{\etal}{{\it et al}}
\newcommand{\CIF}     {\centering \includegraphics[width=2.7in]} %
\newcommand{\Vmin}    {\vspace{-0.2cm}} %
\newcommand{\Hs}      {\hspace{-0.5cm}} %

\doublespace

\begin{document}
\begin{center}
{\Large Deterministic and stochastic algorithms for resolving the flow fields in ducts and networks
using energy minimization}
\par\end{center}{\Large \par}

\begin{center}
Taha Sochi
\par\end{center}

\begin{center}
{\scriptsize University College London, Department of Physics \& Astronomy, Gower Street, London,
WC1E 6BT \\ Email: t.sochi@ucl.ac.uk.}
\par\end{center}

\begin{abstract}
\noindent Several deterministic and stochastic multi-variable global optimization algorithms
(Conjugate Gradient, Nelder-Mead, Quasi-Newton, and Global) are investigated in conjunction with
energy minimization principle to resolve the pressure and volumetric flow rate fields in single
ducts and networks of interconnected ducts. The algorithms are tested with seven types of fluid:
Newtonian, power law, Bingham, Herschel-Bulkley, Ellis, Ree-Eyring and Casson. The results obtained
from all those algorithms for all these types of fluid agree very well with the analytically
derived solutions as obtained from the traditional methods which are based on the conservation
principles and fluid constitutive relations. The results confirm and generalize the findings of our
previous investigations that the energy minimization principle is at the heart of the flow dynamics
systems. The investigation also enriches the methods of Computational Fluid Dynamics for solving
the flow fields in tubes and networks for various types of Newtonian and non-Newtonian fluids.

\vspace{0.3cm}


\par\end{abstract}

\begin{center}

\par\end{center}

\section{Introduction} \label{Introduction}

The traditional method for resolving the pressure and volumetric flow rate fields in fluid
conducting devices is to use the conservation principles, which are normally based on the mass
continuity and momentum conservation, in conjunction with the constitutive relations that link the
stress to the rate of deformation and are specific to the particular types of fluid employed to
model the flow \cite{Skellandbook1967, BirdbookAH1987, Whitebook1991, PapanastasiouGABook1999}. For
single conduits, this usually results in an analytical expression that correlates the volumetric
flow rate to the applied pressure drop as well as other dependencies on the parameters of the
conduits, such as the radius and length of the tube, and the parameters of the fluid such as the
shear viscosity and yield stress. For networks of interconnected conduits, the analytical
expression for the single conduit for the particular fluid model can be exploited in a numeric
solution scheme, which is normally of iterative nature such as the widely used Newton-Raphson
procedure for solving a system of simultaneous non-linear equations, in conjunction with the mass
conservation principle and the given boundary conditions to obtain the flow fields in the network.

Recently the energy minimization principle in the flow through single conduits and networks of
interconnected conduits was investigated \cite{SochiPresSA2014, SochiPresSA22014} as a possible
underlying rule for the flow phenomena that can be exploited to resolve the pressure and volumetric
flow rate fields. While in \cite{SochiPresSA2014} the issue was investigated numerically in
relation to the flow of Newtonian fluids using a stochastic simulated annealing
\cite{MetropolisRRTT1953, KirkpatrickGV1983, Cerny1985} procedure, in \cite{SochiPresSA22014} it
was investigated analytically in relation to the flow of Newtonian and power law fluids using
standard analytical optimization methods from Calculus.

In the present study we continue those investigations but this time the issue is investigated
numerically in relation to the flow of Newtonian and six non-Newtonian fluid models using three
deterministic and one stochastic global multi-variable optimization algorithms. The six
non-Newtonian fluid models are: power law, Bingham, Herschel-Bulkley, Ellis, Ree-Eyring and Casson.
The three employed deterministic algorithms are: Conjugate Gradient, Nelder-Mead, and Quasi-Newton,
while the stochastic algorithm is the Stochastic Global. Several types of network, which include
one-dimensional (1D) two-dimensional (2D) and three-dimensional (3D), of different geometries and
topologies, such as fractals and irregulars based on cubic and orthorhombic lattices, are used to
examine and validate the energy minimization proposal.

All the results obtained in the current study support the generalization of the energy optimization
as a fundamental principle that underlies the flow phenomena in the Newtonian and non-Newtonian
fluid dynamics systems. The study also adds more tools to the Computational Fluid Dynamics as these
computational methods, which are based on energy minimization, can be used for finding the pressure
and flow rate fields in tubes and networks.

The plan for this paper is as follow: in the next section \S\ \ref{TB} we present a general
theoretical background about the energy minimization principle and its use in association with the
global multi-variable optimization algorithms to resolve the flow fields in tubes and networks of
interconnected tubes for different types of fluid. This will be followed in section \S\ \ref{IRA}
by discussing the implementation of the energy minimization principle within global multi-variable
optimization codes and the results that have been obtained from different optimization algorithms
using various types of fluid and different kinds of network where we analyze the results and
compare them to the standard analytical solutions as obtained from and verified by the traditional
methods which are based on the conservation principles and constitutive relations for solving the
flow fields. Finally, in section \S\ \ref{Conclusions} the paper is concluded by outlining the main
issues that have been examined in this study and their theoretical and practical significance.

\section{Theoretical Background}\label{TB}

In this investigation, we assume an incompressible, laminar, pressure-driven, fully-developed flow
with minimal entry and exit effects and negligible viscous frictional losses. We also assume minor
effects from external body forces such as gravitational attraction and electromagnetic interaction.
The single conduits, as well as the conduits in the interconnected networks, are assumed to be
rigid of uniform and circularly-shaped cross sections along their axial dimension.

As for the boundary conditions, we assume Dirichlet-type pressure boundary conditions. The last
assumption is imposed only for convenience and practical considerations; otherwise the energy
minimization argument, when established, will not be restricted to such conditions which basically
reflect the way used to model and portray the flow system by the observer and hence the type of the
boundary conditions does not represent an inherent characteristic of the system that is due to
determine its final outcome.

Concerning the type of fluid, we assume a generalized Newtonian fluid which in this investigation
includes Newtonian, Ostwald-de Waele, Bingham, Herschel-Bulkley, Ellis, Ree-Eyring and Casson
models. The constitutive relations for these models, as well as the analytical expressions for
their volumetric flow rate through rigid uniform pipes of circular cross sections, are given in
Table \ref{QTable}.

The time rate of energy consumption, $I$, for transporting a certain amount of fluid through a
single conducting device, considering the pre-stated flow assumptions, is given by

\begin{equation}
I=\Delta p\, Q
\end{equation}
where $\Delta p$ is the pressure drop across the conducting device and $Q$ is the volumetric flow
rate of the transported fluid through the device. For a flow conducting device that consists of or
discretized into $m$ conducting elements indexed by $l$, the total energy consumption rate,
$I_{t}$, is given by

\begin{equation}\label{ItEq}
I_{t}(p_{1},\ldots,p_{N})=\sum_{l=1}^{m}\Delta p_{l}Q_{l}
\end{equation}
where $N$ is the number of the boundary and internal nodes. For a single duct, the conducting
elements are the discretized sections, while for a network they represent the conducting ducts as
well as their discretized sections if discretization is employed.

Starting from randomly selected values for the internal nodal pressure, with the given pressure
values for the inlet and outlet boundary nodes, the role of the global multi-variable optimization
algorithms in the above-described energy consumption model is to minimize the cost function, which
is the time rate of the total energy consumption for fluid transportation $I_t$ as given by
Equation \ref{ItEq}, by varying the values of the internal nodal pressure while holding the
pressure values at the inlet and outlet boundary nodes as constants. The volumetric flow rates,
$Q$, that have to be used in Equation \ref{ItEq} for the employed fluid models are given by the
expressions in Table \ref{QTable}.

\begin{table} [!h]
\caption{The constitutive relations and the volumetric flow rates, $Q$, for the seven fluid models
used in this investigation. These volumetric flow rates are derived for rigid uniform pipes of
circular cross sections. The meanings of the symbols are given in Nomenclature \S\
\ref{Nomenclature}. \label{QTable}}
\begin{center} 
{\footnotesize
\begin{tabular}{|l|l|l|}
 \hline
Model & \hspace{0.7cm} Constitutive & \hspace{4cm} Q\tabularnewline
 \hline
Newtonian & $\tau=\mu\gamma$ & $\frac{\pi R^{4}\Delta p}{8L\mu}$\tabularnewline
 \hline
Power Law & $\tau=k\gamma^{n}$ & $\frac{\pi R^{4}}{8L}\sqrt[n]{\frac{\Delta
p}{k}}\left(\frac{4n}{3n+1}\right)\left(\frac{2L}{R}\right)^{1-1/n}$\tabularnewline
 \hline
Bingham & $\tau=C\gamma+\tau_{0}$ & $\frac{\pi R^{4}\Delta
p}{8LC}\left[\frac{1}{3}\left(\frac{\tau_{0}}{\tau_{w}}\right)^{4}-\frac{4}{3}\left(\frac{\tau_{0}}{\tau_{w}}\right)+1\right]$\tabularnewline
 \hline
Herschel-Bulkley & $\tau=C\gamma^{n}+\tau_{0}$ & $\frac{8\pi}{\sqrt[n]{C}}\left(\frac{L}{\Delta
p}\right)^{3}\left(\tau_{w}-\tau_{0}\right)^{1+1/n}\left[\frac{\left(\tau_{w}-\tau_{0}\right)^{2}}{3+1/n}+\frac{2\tau_{0}\left(\tau_{w}-\tau_{0}\right)}{2+1/n}+\frac{\tau_{0}^{2}}{1+1/n}\right]$\tabularnewline
 \hline
Ellis &
$\gamma=\frac{\tau}{\mu_{0}}\left[1+\left(\frac{\tau}{\tau_{_{1/2}}}\right)^{\alpha-1}\right]$ &
$\frac{\pi
R^{3}\tau_{w}}{4\mu_{0}}\left[1+\frac{4}{\alpha+3}\left(\frac{\tau_{w}}{\tau_{_{1/2}}}\right)^{\alpha-1}\right]$\tabularnewline
 \hline
Ree-Eyring & $\tau=\tau_{c}\,\mathrm{arcsinh}\left(\frac{\mu_{0}\gamma}{\tau_{c}}\right)$ &
$\frac{\pi
R^{3}\tau_{c}}{\tau_{w}^{3}\mu_{0}}\left[\left(\tau_{c}\tau_{w}^{2}+2\tau_{c}^{3}\right)\cosh\left(\frac{\tau_{w}}{\tau_{c}}\right)-2\tau_{c}^{2}\tau_{w}\sinh\left(\frac{\tau_{w}}{\tau_{c}}\right)-2\tau_{c}^{3}\right]$\tabularnewline
 \hline
Casson & $\tau^{1/2}=\left(K\gamma\right)^{1/2}+\tau_{0}^{1/2}$ & $\frac{\pi
R^{3}}{\tau_{w}^{3}K}\left(\frac{\tau_{w}^{4}}{4}-\frac{4\sqrt{\tau_{0}}\tau_{w}^{7/2}}{7}+\frac{\tau_{0}\tau_{w}^{3}}{3}\right)$\tabularnewline
 \hline
\end{tabular}
}
\end{center}
\end{table}

\section{Implementation, Results and Analysis}\label{IRA}

The above-explained energy minimization method was implemented using three deterministic global
multi-variable optimization algorithms and one stochastic. The deterministic algorithms are:
Conjugate Gradient, Nelder-Mead, and Quasi-Newton, while the stochastic is the Global algorithm of
Boender \etal. For more details about the three employed deterministic algorithms we refer to
standard textbooks that discuss these algorithms such as the Numerical Recipes of Press \etal\
\cite{PressTVF2002}, while for the Stochastic Global algorithm we refer to
\cite{BoenderKTS1982}\footnote{See also: \url{http://jblevins.org/mirror/amiller/global.txt} web
page.}.

\begin{figure}
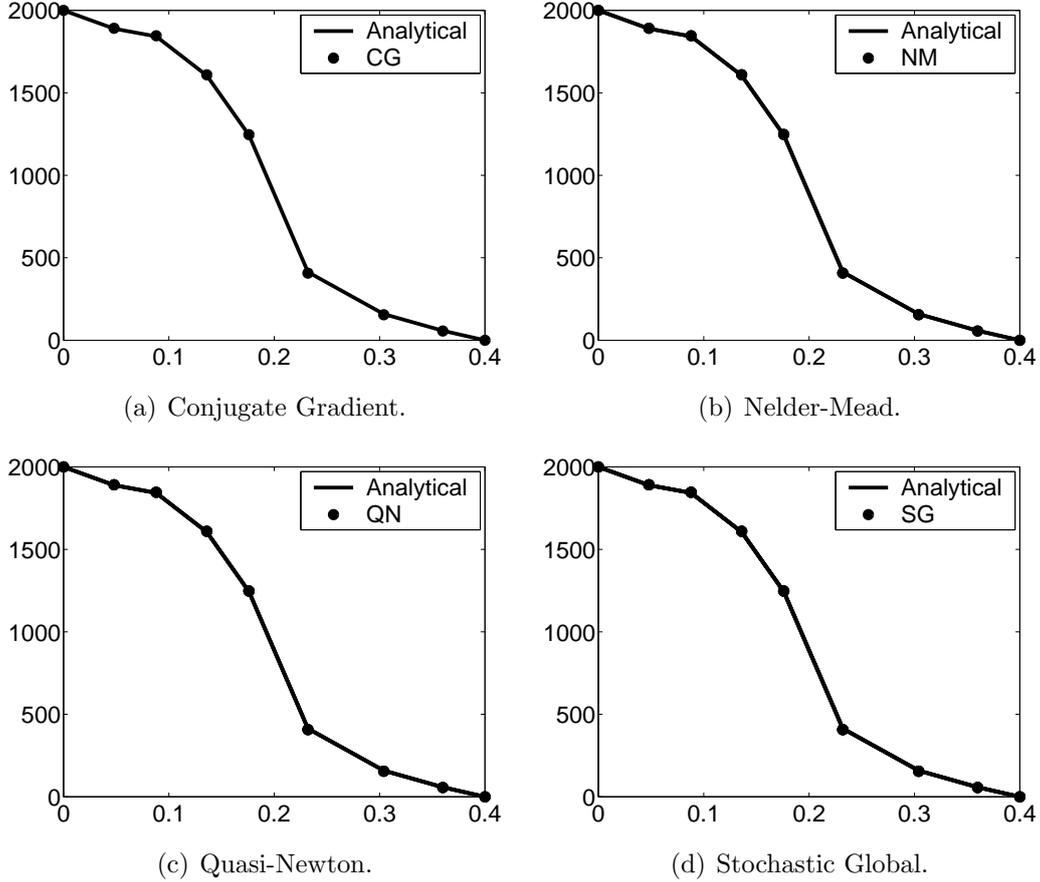

\centering %
\subfigure[Conjugate Gradient.]%
{\begin{minipage}[b]{0.5\textwidth} \CIF {g/HBFig1}
\end{minipage}}
\Hs
\subfigure[Nelder-Mead.]%
{\begin{minipage}[b]{0.5\textwidth} \CIF {g/HBFig2}
\end{minipage}} \Vmin
\centering %
\subfigure[Quasi-Newton.]%
{\begin{minipage}[b]{0.5\textwidth} \CIF {g/HBFig3}
\end{minipage}}
\Hs
\subfigure[Stochastic Global.]%
{\begin{minipage}[b]{0.5\textwidth} \CIF {g/HBFig4}
\end{minipage}}
\caption{Comparison between the analytical solution and the solutions obtained from the indicated
global optimization algorithms which are based on the energy minimization principle for a shear
thinning yield stress Herschel-Bulkley fluid with $n=0.6$, $C=0.008$~Pa.s$^n$ and $\tau_0=1.0$~Pa.
The computations were carried out using the second network of Table \ref{NetTable} with inlet and
outlet pressure boundary conditions of 2000~Pa and 0~Pa respectively. The volumetric flow rate
through the network is $Q=1.022\times 10^{-1}$~m$^3$.s$^{-1}$. In all four sub-figures, the
vertical axis represents the network axial pressure in Pa while the horizontal axis represents the
network axial coordinate in m. Similar results were obtained for the Bingham model which is a
special case of the Herschel-Bulkley model with $n=1$. \label{HBFig}}
\end{figure}

\begin{figure}
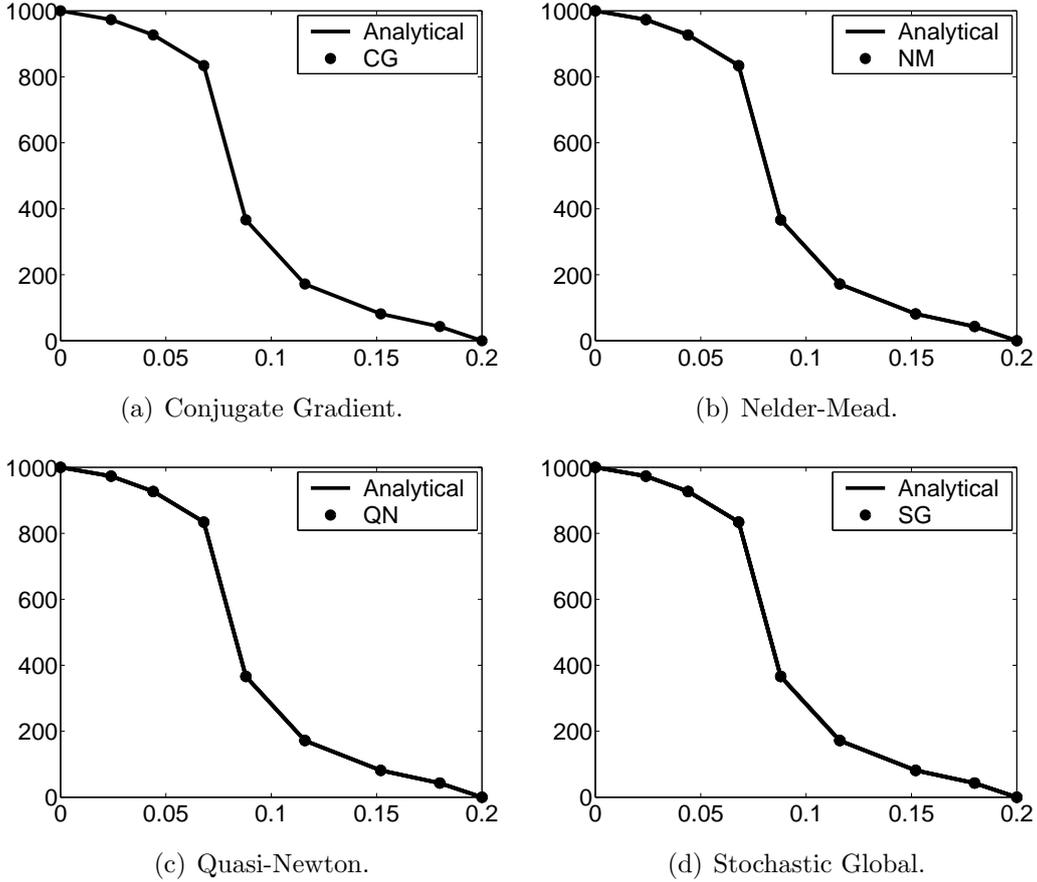

\centering %
\subfigure[Conjugate Gradient.]%
{\begin{minipage}[b]{0.5\textwidth} \CIF {g/EllisFig1}
\end{minipage}}
\Hs
\subfigure[Nelder-Mead.]%
{\begin{minipage}[b]{0.5\textwidth} \CIF {g/EllisFig2}
\end{minipage}} \Vmin
\centering %
\subfigure[Quasi-Newton.]%
{\begin{minipage}[b]{0.5\textwidth} \CIF {g/EllisFig3}
\end{minipage}}
\Hs
\subfigure[Stochastic Global.]%
{\begin{minipage}[b]{0.5\textwidth} \CIF {g/EllisFig4}
\end{minipage}}
\caption{Comparison between the analytical solution and the solutions obtained from the indicated
global optimization algorithms which are based on the energy minimization principle for an Ellis
fluid with $\mu_0=0.18$~Pa.s, $\alpha=2.4$ and $\tau_{_{1/2}}=1025$~Pa. The computations were
carried out using the third network of Table \ref{NetTable} with inlet and outlet pressure boundary
conditions of 1000~Pa and 0~Pa respectively. The volumetric flow rate through the network is
$Q=3.160\times 10^{-6}$~m$^3$.s$^{-1}$. In all four sub-figures, the vertical axis represents the
network axial pressure in Pa while the horizontal axis represents the network axial coordinate in
m. \label{EllisFig}}
\end{figure}

As for single tubes, it is a special case of the forthcoming linear networks of serially connected
pipes where all the pipes in the ensemble have the same radius; in this regard all the implemented
optimization algorithms produced results which are virtually identical to the analytical solutions
for all the seven types of fluid as given in Table \ref{QTable}. Regarding the networks, due to the
difficulty of presenting the results graphically for the two-dimensional and three-dimensional
networks, we present in Figures \ref{PLFig}--\ref{CasFig} a sample of the results obtained from a
range of one-dimensional networks presented in Table \ref{NetTable}. Similar results were obtained
from representative samples of two-dimensional and three-dimensional networks although the
numerical errors for the two-dimensional and three-dimensional networks are generally larger than
those of the one-dimensional. Also some algorithms failed to converge in the case of large networks
due to shortcomings of the employed algorithms and codes or restrictions on the affordable CPU time
or the number of iterations of their execution.

\begin{figure}
\centering %
\subfigure[One-dimensional Linear Network.]%
{\begin{minipage}[b]{0.5\textwidth} \centering \includegraphics[height=0.8cm, width=9cm]
{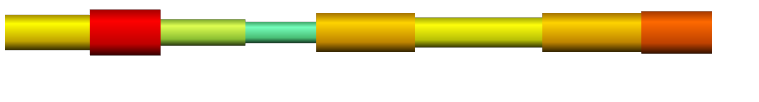}
\end{minipage}} \vspace{-0.3cm}
\centering %
\subfigure[Two-dimensional Fractal Network.]%
{\begin{minipage}[b]{0.5\textwidth} \centering \includegraphics[height=9cm, width=9cm]
{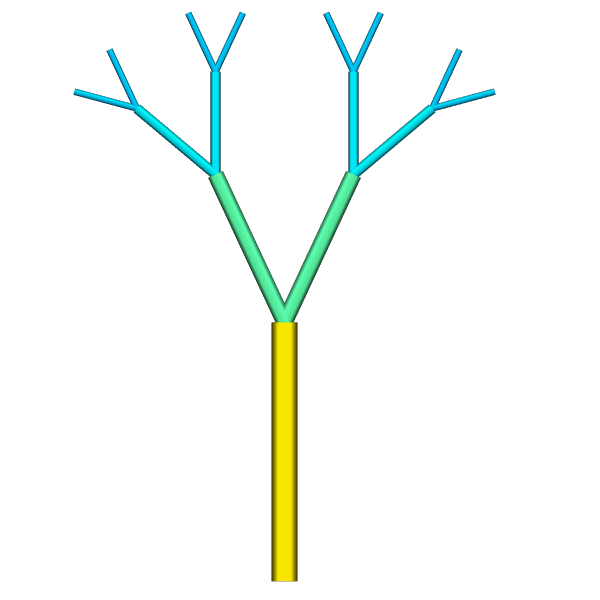}
\end{minipage}} \vspace{-0.3cm}
\centering %
\subfigure[Three-dimensional Orthorhombic Network.]%
{\begin{minipage}[b]{0.5\textwidth} \centering \includegraphics[height=6cm, width=9cm]
{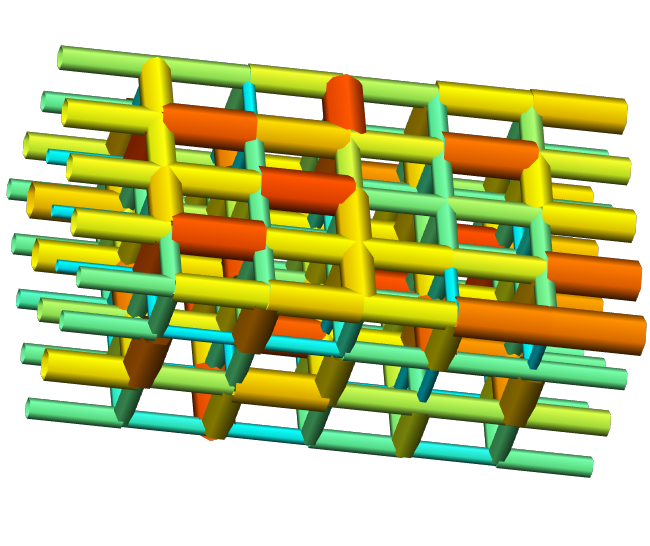}
\end{minipage}}
\caption{A graphic demonstration of three main types of network used in the current investigation.
\label{GraphicNets}}
\end{figure}

The size of the networks used in the investigation ranges between a small number to several
hundreds, and even thousands in some cases, of nodes and segments. The multi-dimensional networks
used in this investigation are of two main types: two-dimensional of fractal and rectangular
morphology, and three-dimensional built on cubic and orthorhombic lattice structures. The fractals
are based on fractal branching patterns where each generation of the branching tubes in the network
has a specific number of branches related to the number of branches in the parent generation, such
as 3:1, as well as specific branching angle, radius branching ratio and length to radius ratio. The
cubic and orthorhombic networks are based on cubic and orthorhombic three-dimensional lattice
structures respectively where the radii of the tubes in the network are subjected to random
statistical distributions such as the uniform or the normal distributions. Similar statistical
distributions were also applied to the two-dimensional rectangular networks. A graphic
demonstration of three main types of network; namely one-dimensional linear, two-dimensional
fractal and three-dimensional orthorhombic; is given in Figure \ref{GraphicNets}.


The size of the difference between the numerical optimization solutions and the analytical
solutions depends mainly on the particular algorithm, the type and parameters of the fluid and the
size and type (1D, 2D or 3D and fractal or orthorhombic) of the network. A typical size of the
average percentage relative difference between the numerical optimization solutions and the
analytical solutions is less than 0.5\% for the one-dimensional networks, about 1\% for the
two-dimensional networks, and 2-3\% for the three-dimensional networks. In most cases, the best
optimization algorithm with regard to the agreement of its solution with the analytical solution is
the Global while the worst is the Nelder-Mead. The latter has also failed to converge in some
cases.

In our view, the observed discrepancy between the numerical optimization solutions and the
analytical solutions in all the investigated cases can be justified by premature convergence of the
optimization algorithms due to practical limits on their convergence criteria as well as numerical
errors arising from limitations of the employed optimization algorithms and codes plus
non-linearities, especially in some cases of non-Newtonian models with extreme non-linear
characteristics such as high shear thinning and yield stress.

There are many computational issues related to the performance and convergence behavior of these
algorithms and their relation to the type and parameters of the fluids and the size and type of the
networks. However, these technical details are irrelevant to the current study whose main objective
is to provide further validation and demonstration for the use of energy minimization principle in
resolving the flow fields in tubes and networks, rather than investigating numerical and
computational issues.


\begin{table} [!h]
\caption{A sample of the one-dimensional linear networks of serially connected rigid uniform tubes
of circular cross sections with the given number of segments (NS) that have been used in the
current investigation to compare the analytical solutions to the solutions of the global
optimization algorithms. \label{NetTable}}
\begin{center} 
\begin{tabular}{c|c|c|c}
\hline
{\bf Network} &   {\bf NS} & {\bf Lengths (cm)} & {\bf Radii (cm)} \\
\hline
   {\bf 1} &          7 & 80,60,70,90,90,50,60 & 2.5,2.1,1.8,1.3,1.7,2.6,1.6 \\

   {\bf 2} &          8 & 4.8,4,4.8,4,5.6,7.2,5.6,4 & 1,1.3,0.75,0.6,0.5,0.85,1.1,1.2 \\

   {\bf 3} &          8 & 2.4,2,2.4,2,2.8,3.6,2.8,2 & 0.6,0.5,0.44,0.28,0.38,0.49,0.57,0.51 \\

   {\bf 4} &          6 & 14,8,8,14,14,17 & 2.4,2,1.76,2.232,1.52,1.8 \\
\hline
\end{tabular}
\end{center}
\end{table}


\begin{figure}
\centering %
\subfigure[Conjugate Gradient.]%
{\begin{minipage}[b]{0.5\textwidth} \CIF {g/PLFig1}
\end{minipage}}
\Hs
\subfigure[Nelder-Mead.]%
{\begin{minipage}[b]{0.5\textwidth} \CIF {g/PLFig2}
\end{minipage}} \Vmin
\centering %
\subfigure[Quasi-Newton.]%
{\begin{minipage}[b]{0.5\textwidth} \CIF {g/PLFig3}
\end{minipage}}
\Hs
\subfigure[Stochastic Global.]%
{\begin{minipage}[b]{0.5\textwidth} \CIF {g/PLFig4}
\end{minipage}}
\caption{Comparison between the analytical solution and the solutions obtained from the indicated
global optimization algorithms which are based on the energy minimization principle for a shear
thickening power law fluid with $n=1.2$ and $k=0.05$~Pa.s$^n$. The computations were carried out
using the first network of Table \ref{NetTable} with inlet and outlet pressure boundary conditions
of 3000~Pa and 0~Pa respectively. The volumetric flow rate through the network is $Q=1.690\times
10^{-4}$~m$^3$.s$^{-1}$. In all four sub-figures, the vertical axis represents the network axial
pressure in Pa while the horizontal axis represents the network axial coordinate in m. Similar
results were obtained for the Newtonian model which is a special case of the power law model with
$n=1$. \label{PLFig}}
\end{figure}


\begin{figure}
\centering %
\subfigure[Conjugate Gradient.]%
{\begin{minipage}[b]{0.5\textwidth} \CIF {g/REFig1}
\end{minipage}}
\Hs
\subfigure[Nelder-Mead.]%
{\begin{minipage}[b]{0.5\textwidth} \CIF {g/REFig2}
\end{minipage}} \Vmin
\centering %
\subfigure[Quasi-Newton.]%
{\begin{minipage}[b]{0.5\textwidth} \CIF {g/REFig3}
\end{minipage}}
\Hs
\subfigure[Stochastic Global.]%
{\begin{minipage}[b]{0.5\textwidth} \CIF {g/REFig4}
\end{minipage}}
\caption{Comparison between the analytical solution and the solutions obtained from the indicated
global optimization algorithms which are based on the energy minimization principle for a
Ree-Eyring fluid with $\mu_0=0.018$~Pa.s and $\tau_{c}=300$~Pa. The computations were carried out
using the fourth network of Table \ref{NetTable} with inlet and outlet pressure boundary conditions
of 1500~Pa and 0~Pa respectively. The volumetric flow rate through the network is $Q=4.991\times
10^{-3}$~m$^3$.s$^{-1}$. In all four sub-figures, the vertical axis represents the network axial
pressure in Pa while the horizontal axis represents the network axial coordinate in m.
\label{REFig}}
\end{figure}


\begin{figure}
\centering %
\subfigure[Conjugate Gradient.]%
{\begin{minipage}[b]{0.5\textwidth} \CIF {g/CasFig1}
\end{minipage}}
\Hs
\subfigure[Nelder-Mead.]%
{\begin{minipage}[b]{0.5\textwidth} \CIF {g/CasFig2}
\end{minipage}} \Vmin
\centering %
\subfigure[Quasi-Newton.]%
{\begin{minipage}[b]{0.5\textwidth} \CIF {g/CasFig3}
\end{minipage}}
\Hs
\subfigure[Stochastic Global.]%
{\begin{minipage}[b]{0.5\textwidth} \CIF {g/CasFig4}
\end{minipage}}
\caption{Comparison between the analytical solution and the solutions obtained from the indicated
global optimization algorithms which are based on the energy minimization principle for a Casson
fluid with $\mu_0=0.01$~Pa.s and $\tau_{0}=1.0$~Pa. The computations were carried out using the
fourth network of Table \ref{NetTable} with inlet and outlet pressure boundary conditions of
2500~Pa and 0~Pa respectively. The volumetric flow rate through the network is $Q=9.627\times
10^{-3}$~m$^3$.s$^{-1}$. In all four sub-figures, the vertical axis represents the network axial
pressure in Pa while the horizontal axis represents the network axial coordinate in m.
\label{CasFig}}
\end{figure}


\clearpage
\section{Conclusions} \label{Conclusions}

In this study, energy minimization was examined as a principle that underlies the flow phenomena in
single tubes and networks of interconnected tubes. This was demonstrated by using three
deterministic (Conjugate Gradient, Nelder-Mead and Quasi-Newton) and one stochastic (Global)
multi-variable global optimization algorithms. Seven fluid models (Newtonian, power law, Bingham,
Herschel-Bulkley, Ellis, Ree-Eyring and Casson) with different types of network (1D linear, 2D
fractal, 2D rectangular, 3D cubic and 3D orthorhombic) were used in this investigation. All the
obtained results support the validity and generality of the energy minimization principle. The
outcome of this investigation lends more credibility to the previous findings in
\cite{SochiPresSA2014, SochiPresSA22014} about this issue; moreover it generalizes the validity of
the principle by extending its applicability to more types of fluid which include several widely
used non-Newtonian models.

Apart from the obvious theoretical significance of the findings of the previous and current
investigations, the optimization algorithms can be used to resolve the pressure and volumetric flow
rate fields. Although these algorithms may not be the best in performance, and even in accuracy in
some cases, they could have practical applications in the case of very large networks where the use
of the traditional methods, which are based on the conservation principles and constitutive
relations, is prohibitive due to the requirement of using very large matrices. The role and
justification of the use of the optimization algorithms in resolving the flow fields is similar to
the role and justification of their use in solving large combinatorial problems, like the Traveling
Salesman Problem, where other analytical or conceptually-based methods that rely on direct
combinatorial enumeration are not viable or available in those circumstances.

\clearpage
\section{Nomenclature} \label{Nomenclature}

\begin{supertabular}{ll}
$\alpha$                &   indicial parameter in Ellis model \\
$\gamma$                &   rate of shear strain \\
$\mu$                   &   Newtonian viscosity \\
$\mu_0$                 &   low-shear viscosity in Ellis and Ree-Eyring models \\
$\tau$                  &   shear stress \\
$\tau_0$                &   yield stress in Herschel-Bulkley and Casson models \\
$\tau_{_{1/2}}$         &   shear stress when the viscosity equals $\frac{\mu_0}{2}$ in Ellis model \\
$\tau_c$                &   characteristic shear stress in Ree-Eyring model \\
$\tau_{w}$              &   shear stress at tube wall $(=\frac{R \Delta p}{2L})$ \\
\\
$C$                     &   viscosity consistency coefficient in Bingham and Herschel-Bulkley models \\
$I$                     &   time rate of energy consumption for fluid transport \\
$I_t$                   &   time rate of total energy consumption for fluid transport \\
$k$                     &   viscosity consistency coefficient in power law model \\
$K$                     &   viscosity consistency coefficient in Casson model \\
$L$                     &   tube length \\
$m$                     &   number of discrete elements in the fluid conducting device \\
$n$                     &   index of power law and Herschel-Bulkley models \\
$N$                     &   number of nodal junctions in the fluid conducting device \\
$p$                     &   pressure \\
$\Delta p$              &   pressure drop across flow conduit \\
$Q$                     &   volumetric flow rate \\
$R$                     &   tube radius \\
\end{supertabular}

\newpage
\phantomsection \addcontentsline{toc}{section}{References} %
\bibliographystyle{unsrt}

\end{document}